\documentclass[
superscriptaddress,
reprint,preprintnumbers,
amsmath,amssymb,
aps,
nofootinbib,
nobibnotes]{revtex4-1}

\usepackage{graphicx}
\usepackage{dcolumn}
\usepackage{bm}
\usepackage{slashed}
\usepackage{ulem}
\usepackage{lipsum}
\usepackage{hyperref}
\usepackage{multirow}
\usepackage{subfig}
\usepackage[usenames]{color}

\hypersetup{
	colorlinks = true,
	linkcolor = blue,
	citecolor = red
}

\def\lsim{\mathrel{\rlap{\lower4pt\hbox{\hskip1pt$\sim$}}
    \raise1pt\hbox{$<$}}}
    
\def\gsim{\mathrel{\rlap{\lower4pt\hbox{\hskip1pt$\sim$}}
    \raise1pt\hbox{$>$}}}

\begin{document}

\title{Search for sterile neutrino with light gauge interactions: \\  
recasting collider, beam-dump, and neutrino telescope searches}

\author{Yongsoo Jho}
\email{jys34@yonsei.ac.kr}
\affiliation{Department of Physics and IPAP, Yonsei University, Seoul 03722, Republic of Korea}

\author{Jongkuk Kim}
\email{jkkim@kias.re.kr}
\affiliation{Korea Institute for Advanced Study, Seoul 02455, Republic of Korea}

\author{Pyungwon Ko}
\email{pko@kias.re.kr}
\affiliation{Korea Institute for Advanced Study, Seoul 02455, Republic of Korea}

\author{Seong Chan Park}
\email{sc.park@yonsei.ac.kr}
\affiliation{Department of Physics and IPAP, Yonsei University, Seoul 03722, Republic of Korea}

\preprint{KIAS-P20046, LDU-2020-07}
\date{\today}

\begin{abstract}
We investigate features of the sterile neutrinos in the presence of a light gauge boson $X^\mu$ that couples to the neutrino sector. The novel bounds on the active-sterile neutrino mixings $| U_{\ell 4} |^2$, especially for tau flavor ($l = \tau$), from various collider and fixed target experiments are explored.  Also, taking into account the additional decay channel of the sterile neutrino  into a light gauge boson ($\nu_4 \to 
\nu_\ell e^+ e^-$), 
we explore and constrain a parameter space for low energy excess in neutrino oscillation experiments.
\end{abstract}


\maketitle

\section{Introduction}

The sterile neutrinos having no known non-gravitational couplings with the standard model (SM) 
particles have been seriously considered to interpret the recent observational anomalies in the neutrino oscillations, such as Low Energy Excess (LEE) reported from MiniBooNE~\cite{AguilarArevalo:2007it, AguilarArevalo:2008rc, AguilarArevalo:2010wv, Aguilar-Arevalo:2013pmq, Aguilar-Arevalo:2018gpe, Aguilar-Arevalo:2020nvw} and LSND \cite{Athanassopoulos:1995iw, Athanassopoulos:1996jb, Athanassopoulos:1997pv, Athanassopoulos:1996wc} experiments.  They are also important targets to be discovered in fixed target and collider experiments~\cite{Ilten:2018crw, Bauer:2018onh} when sterile neutrinos are long-lived. 

Since all the fermions in the SM have gauge interactions, it may not be surprising if the sterile neutrinos are actually charged under some gauge symmetries. As minimal attempts to incorporate the gauge interactions of the sterile neutrinos, we consider two theoretically well motivated models with additional $U(1)_X$ symmetry extending the SM gauge group~\footnote{The most generic anomaly free extension of the SM  is to gauge the difference of lepton numbers as well: $(B_k-L_k) + \epsilon(L_i-L_j)$ where $i,j,k =1,2,3$ generations~\cite{Foot:1990mn, He:1990pn,He:1991qd}. Also see \cite{Chun:2018ibr, Ko:2019tts, Jho:2020jsa}.}
:
\begin{itemize} 
\item[(i)] The $U(1)_{B-L}$ symmetry of the baryon($B$) and lepton($L$) numbers~\cite{Marshak:1979fm, Mohapatra:1980qe, Davidson:1978pm} and the sterile neutrinos are identified as the right-handed neutrinos carrying the lepton number $L = 1$. 
\item[(ii)] The sterile-neutrino-specific $U(1)_s$ symmetry and the sterile neutrino has the charge $Q_s=1$.  All the SM fermions are singlets~\cite{Ballett:2018ynz, Bertuzzo:2018itn} 
\end{itemize}
 In the two models, the lightest new non-active neutrino component is commonly called the `the sterile neutrino' and is denoted as $\nu_4$ in mass eigenstate. The interactions of $\nu_4$ with the SM fermions depend on the sterile-active mixings among neutrinos and also the kinetic mixing of the new gauge boson and the hypercharge gauge boson (or photon below the electroweak scale). Due to the new gauge interactions, the conventional search do not efficiently cover the parameter space of the models and call for new strategies. Therefore, our goal is to provide the theoretical framework to define the proper parameters of the models and then provide the comprehensive study on the experimental bounds from existing searches and also foresee the future expectations.

The paper is organized as follows: The two gauge invariant, realistic UV complete models are introduced in Sec.~\ref{sec:model} then the decay width (and the length) of $\nu_4$ is calculated in Sec.~\ref{sec:width}. The Sec.~\ref{sec:constraints} is devoted to the various experiments from beam-dump experiments such as CHARM \cite{Orloff:2002de} and NOMAD \cite{Astier:2001ck}, the double-cascade search at the IceCube with up-going atmospheric neutrinos \cite{Coloma:2017ppo}, as well as the collider searches from Belle/BaBar \cite{Kobach:2014hea, Kim:2019xqj} and LEP-I \cite{Abreu:1996pa}. We evaluate the expected sensitivities on the active-sterile mixing $| U_{\ell 4}|^2$ from the long-lived particle searches at FASER/FASER2  \cite{Ariga:2018uku} and SHiP \cite{Alekhin:2015byh} especially for $\ell=\tau$.  We conclude our study in Sec.~\ref{sec:conclusion} suggesting future long-lived particle searches at the intensity frontier can be remarkably useful to probe intriguing new physics scenarios to explain the low energy excess in the electron spectrum in neutrino oscillation measurement.



\section{Model}\label{sec:model}

\subsection{Model A: Gauged $U(1)_{B-L}$}

An  extension of the SM gauge group as $G = G_{\rm SM} \times U(1)_X$ with $X = B-L$ is introduced 
for Model A. There should be three additional right-handed neutrinos $\hat{N}_{iR}$ ($i=1,2,3$) 
to cancel out the $U(1)_{B-L}$ anomaly~\cite{Marshak:1979fm, Mohapatra:1980qe, Davidson:1978pm, Buchmuller:1991ce, Heeba:2019jho, Mohapatra:2019ysk, Mohapatra:2020bze}. The matter content in the model is shown in Table.~\ref{modelA_Field_Content}.
\begin{table}[h]
\begin{tabular}{| c | c | c | c | c | c | c |} 
\hline
& \ \ $SU(3)_c$ \ \ & \ \ $SU(2)_L$ \ \ & \ \ \ $Y$ \ \ \ & \ \ $B-L$ \ \ \\
\hline
$\hat{Q}_i = (\hat{u}_i \ \hat{d}_i)^T$ & {\bf 3} & {\bf 2} & $+\frac{1}{6}$ & $+\frac{1}{3}$ \\
$\hat{u}_i^c$ & $\bar{{\bf 3}}$ & {\bf 1} & $-\frac{2}{3}$ & $-\frac{1}{3}$ \\
$\hat{d}_i^c$ & $\bar{{\bf 3}}$ & {\bf 1} & $+\frac{1}{3}$ & $-\frac{1}{3}$ \\
\hline
$\hat{L}_i = (\hat{\nu}_i \ \hat{e}_i)^T$ & {\bf 1} & {\bf 2} & $-\frac{1}{2}$ & $-1$ \\
$\hat{e}_i^c$ & {\bf 1} & {\bf 1} & $+1$ & $+1$ \\
$\hat{N}_{4,5,6}^c$ & {\bf 1} & {\bf 1} & $0$ & $+1$ \\
\hline
$H$ & ${\bf 1}$ & {\bf 2} & $+\frac{1}{2}$ & 0 \\
$\Phi$ & {\bf 1} & {\bf 1} & 0 & $+2$  \\
\hline
\end{tabular}
    \caption{(Model A) Matter field content in interaction basis. All fermions are left-handed. The index $i=1,2,3$ is for three generations of fermions. For SM LH neutrinos, $\hat{\nu}_1 = \nu_e$, $\hat{\nu}_2 = \nu_\mu$ and $\hat{\nu}_3 = \nu_\tau$.}
    \label{modelA_Field_Content}   
\end{table}

The most general gauge invariant Lagrangian is written in terms of the interaction eigenstates as
\begin{eqnarray}
\mathcal{L} & = & \mathcal{L}_{\rm SM} - \frac{1}{4} X_{\mu \nu} X^{\mu \nu} \nonumber \\
& & + \frac{1}{2} | (\partial_\mu + 2 i g_X X_\mu ) \Phi |^2 - \frac{\lambda_\Phi}{4} \Bigl ( | \Phi |^2 - v_\Phi^2 \Bigr )^2 \nonumber \\ 
& & - \frac{g_X}{3} \Bigl ( \overline{\hat{Q}_{iL}} \gamma^\mu \hat{Q}_{iL} + \overline{\hat{u}_{iR}} \gamma^\mu \hat{u}_{iR} + \overline{\hat{d}_{iR}} \gamma^\mu \hat{d}_{iR} \Bigr ) X_\mu \nonumber \\
& & + g_X \Bigl (\overline{\hat{L}_{i L}} \gamma^\mu \hat{L}_{i L} + \overline{\hat{e}_{i R}} \gamma^\mu \hat{e}_{i R} + \overline{\hat{N}_{j'R}} \gamma^\mu \hat{N}_{j'R} \Bigr ) X_\mu \nonumber \\
& & - \Bigl( \hat{y}_\nu^{ii'} \overline{\hat{L}_{iL}} \tilde{H} \hat{N}_{i'R} + \frac{\hat{\lambda}_R^{i'j'}}{2} \Phi \overline{\hat{N}_{i'R}^c} \hat{N}_{j'R} + \text{h.c.} \Bigr ),
\end{eqnarray}
where $i=1,2,3$, and $i',j'=4,5,6$.
The vacuum expectation value (VEV) of $\Phi$ is non-vanishing for the spontaneous symmetry breaking of $U(1)_X$. The mixings between six Majorana neutrinos are parametrized by the PMNS matrix elements $U_{\ell i}^{\rm PMNS}, i=1,2,3$ and also by $U_{\ell j'}, j'=4,5,6$ for active-sterile mixings:
\begin{eqnarray}
\nu_{\ell L} & = & \sum_{i=1,2,3} U_{\ell i}^{\rm PMNS} \nu_{i L} + \sum_{j'=4,5,6} U_{\ell j'} N_{j' L}^c \ \ \ (\ell =e,\mu,\tau), \ \ \ \ \ 
\end{eqnarray}

Because of the non-trivial active-sterile mixings, the neutrinos interact with the $X$ boson (up to the linear order of $U_{\tau 4}$). In particular, we note that the $\nu_4$-$\nu_l$-$X$ interaction are allowed especially for $\ell=\tau$: 
\begin{eqnarray}
\mathcal{L}_{\nu\text{-}X} & \supset & + g_X U_{\tau 4} \bar{\nu}_\tau \gamma^\mu P_L \nu_4 X_\mu + \mathcal{O}(|U_{\tau4}|^2).
\end{eqnarray}
In model A, we focus on the case of $m_4 > m_X$ and  the total width of $\nu_4$ is greatly enhanced by the new gauge interactions even with a small gauge coupling constant, $g_X \sim 10^{-6} - 10^{-4}$.

\subsection{Model B: Sterile $\nu$-specific $U(1)_s$}

As our second choice of the model, we consider a dark `sterile neutrino-specific' gauge interaction with a hidden $U(1)_s$ ($s=X$) symmetry under which no SM particles are charged 
\footnote{ In case dark matter 
of $\sim O(1)$ TeV is also charged under the same $U(1)_s$ and the sterile neutrino is $\sim O(1)$ eV,
then some puzzles in the CDM paradigm can be resolved by strong self interactions between DM
\cite{Bringmann:2013vra},  and $\nu_s$ can behave as dark radiation in the early universe \cite{Ko:2014bka}.}. 
The particle content and the charge assignment of Model B is shown in Table.~\ref{modelB_Field_Content}.

\begin{table}[t]
\begin{tabular}{| c | c | c | c | c | c | c |} 
\hline
& \ \ $SU(3)_c$ \ \ & \ \ $SU(2)_L$ \ \ & \ \ \ $Y$ \ \ \ & \ \ $s$ \ \ \\
\hline
$\hat{Q}_i = (\hat{u}_i \ \hat{d}_i)^T$ & {\bf 3} & {\bf 2} & $+\frac{1}{6}$ & $0$ \\
$\hat{u}_i^c$ & $\bar{{\bf 3}}$ & {\bf 1} & $-\frac{2}{3}$ & $0$ \\
$\hat{d}_i^c$ & $\bar{{\bf 3}}$ & {\bf 1} & $+\frac{1}{3}$ & $0$ \\
\hline
$\hat{L}_i = (\hat{\nu}_i \ \hat{e}_i)^T$ & {\bf 1} & {\bf 2} & $-\frac{1}{2}$ & $0$ \\
$\hat{e}_i^c$ & {\bf 1} & {\bf 1} & $+1$ & $0$ \\
 $\hat{N}_R$ & {\bf 1} & {\bf 1} & $0$ & $0$  \\
$\hat{\nu}_{s}$ & {\bf 1} & {\bf 1} & $0$ & $+1$ \\
\hline
$H$ & ${\bf 1}$ & {\bf 2} & $+\frac{1}{2}$ & 0 \\
$\phi$ & {\bf 1} & {\bf 1} & 0 & $+1$  \\
\hline
\end{tabular}
    \caption{ (Model B) Matter field content in interaction basis. All SM fermions are left-handed. The index $i=1,2,3$ is for three generations of fermions.
   }
    \label{modelB_Field_Content}   
\end{table}

Having introduced gauge singlet RH neutrinos $N_R$'s, one $U(1)_s$-charged SM-singlet 
Dirac neutrino $\hat{\nu}_s$ and also new scalar field, $\phi$,  in Model B, the Lagrangian of the Model is given as \cite{Ko:2014bka} 
\begin{align}
\mathcal{L} & =  
\mathcal{L}_{\rm SM} - \frac{1}{4} X_{\mu \nu} X^{\mu \nu} + \frac{m_X^2}{2} X_\mu X^\mu \nonumber \\
&  - \frac{\epsilon_{\gamma X}}{2} X_{\mu \nu} B^{\mu \nu} 
- g_X \bar{\nu}_s \gamma^\mu \nu_s X_\mu - m_s \overline{\hat{\nu}_{s}} \hat{\nu}_{s}
\nonumber\\
&-\left( \hat{y}_{\nu_s}^{j} \overline{\hat{N}_{Rj}} \hat{\nu}_{s} \phi^\dagger + \hat{y}_\nu^{ij}\overline{\hat{L}_i} \tilde{H} \hat{N}_{Rj} + {\text h.c.} \right).
\end{align}
The newly introduced scalar $\phi$ is the SM singlet but have $U(1)_X$ charge $Q_X(\phi) = +1$. Non-vanishing VEV of the new Higgs leads to the mixing between $\hat{\nu}_s$ and $\hat{N}_R$ neutrinos. From the Dirac mass mixing, we read $U_{\ell 4} \sim \hat{y}_{\nu_s} v_\phi /m_s$. Taking active neutrino mass as $m_\nu \sim 0.1$eV, we obtain $\hat{y}_\nu \sim 10^{-12}.$ Depending on a value of $\hat{y}_{\nu_s}$ and $m_s$, the mixing angle $U_{\ell  4}$ is determined. 

Just as Model A, in the mass eigenbasis, $X$ boson couples to the active neutrinos as well as the charged SM fermions via active-sterile mixing and the kinetic mixing: 
\begin{eqnarray}
\mathcal{L} & \supset & - g_X U_{\ell 4} \bar{\nu}_\ell \gamma^\mu \nu_4 X_\mu - g_X \epsilon_{\gamma X} \cos \theta_W Q_f \bar{f} \gamma^\mu f X_\mu, \ \ \ \label{modelB_nu_int}
\end{eqnarray}
where $\nu_4$ is a nearly sterile neutrino in the mass basis, after diagonalization. 
Considering the original coupling $g_X$ of order 1 and the kinetic mixing between $X$ and SM hypercharge boson as $\epsilon_{\gamma X} \sim \mathcal{O}(10^{-4}-10^{-3})$, the model containing interactions of Eq.~\ref{modelB_nu_int} has been suggested to interpret MiniBooNE/LSND low energy excess with $m_X$, $m_4$ in the range of MeV-GeV \cite{Ballett:2018ynz, Bertuzzo:2018itn}. 
The best-fit scenario with $|U_{\mu4}|^2 \sim 10^{-7} - 10^{-6}$ and/or $| U_{\tau4} |^2 \sim 10^{-4}$ 
are expected in the literatures, which can be probed by various long-lived particle searches and rare 
meson decays \footnote{Work in progress.}. For model B, when $g_X \sim 1$, the prompt two body 
decay $\nu_4 \to \nu_\ell + X$ is allowed unless $m_4 <m_X$. However, even in this case, 
the three-body decay $\nu_s \to \nu_\ell + \ell \bar{\ell}$ becomes important as we will study 
in detail in the next section.


\section{Total width and decay length of sterile neutrino}
\label{sec:width}

\subsection{Conventional CC/NC decays of sterile neutrino}

Even no $U(1)_X$ interactions in Model A and Model B are included, the $\nu_4$ decays to neutrinos, electron and muon and also to neutral pseudocscalar mesons ($P^0=\pi^0, \eta, \eta'$) and vector mesons ($V^0=\rho, \omega, \phi$) by the SM charged current (CC) and neutral current (NC) interactions through the $|U_{\tau4}|^2$ mixing  \cite{Helo:2010cw}: 
\begin{align}
\Gamma_{\nu_4,\text{total}}^{\text{(NC/CC)}} 
&= \sum_{\ell=e,\mu,\tau} \Gamma (\nu_4 \to \nu_\tau \nu_\ell \bar{\nu}_\ell) 
+ \sum_{l=e,\mu} \Gamma (\nu_4 \to \nu_\tau \ell^- \ell^+) \nonumber \\
&+  \sum_{P=\pi,\eta,\eta', V=\rho,\omega,\phi} \Gamma(\nu_4 \to \nu_\tau P^0/V^0) 
\end{align}
where 
\begin{eqnarray}
\Gamma (\nu_4 \rightarrow \nu_\tau P^0) & = & \frac{G_F^2}{64\pi} f_P^2 | U_{\tau4} |^2 m_4^3 \ (1 - x_P^2)^2, \\
\Gamma (\nu_4 \rightarrow \nu_\tau V^0) & = & \frac{G_F^2}{2\pi} \kappa_V^2 f_V^2 | U_{\tau 4} |^2 m_4^3 \nonumber \\
& & \times (1- x_V^2)^2 (1+2 x_V^2), \\
\Gamma (\nu_4 \rightarrow \ell_1^- \ell_1^+ \nu_\tau) & = & \frac{G_F^2}{96\pi^3} | U_{\tau 4} |^2 m_4^5 \ \Bigl [ g_L^{\ell_1} g_R^{\ell_1} I_2 (x_{\nu_\tau} , x_{\ell_1}, x_{\ell_1}) \nonumber \\
& & + \Bigl ( (g_L^{\ell_1})^2 + (g_R^{\ell_1})^2 \Bigr ) I_1 (x_{\nu_\tau}, x_{\ell_1}, x_{\ell_1}) \Bigr ], \nonumber \\
& & \\
\Gamma(\nu_4 \rightarrow \nu_\tau \nu_{\ell_2} \bar{\nu}_{\ell_2} ) & = & \frac{G_F^2}{96\pi^3} | U_{\tau 4} |^2 m_4^5.
\end{eqnarray}
The functions $I_1 (x,y,z)$, $I_2 (x,y,z)$ are given in Ref.~\cite{Helo:2010cw}. 
The leptonic couplings are $g_L^l = - \frac{1}{2} + \sin^2 \theta_W$, $g_R^l = \sin^2 \theta_W$ and the NC coupling to neutral mesons are $\kappa_V^{\rho, \omega} = \frac{1}{3} \sin^2 \theta_W$, $\kappa_V^\phi = - \frac{1}{4} + \frac{1}{3} \sin^2 \theta_W$. $f_{P,V}$ are the decay constant of each pseudoscalar ($P$) and vector ($V$) mesons \cite{Helo:2010cw}.

\begin{figure}[t]
\centering
\subfloat[$\nu_4 \rightarrow \nu_\tau X$]
{\includegraphics[width=0.225\textwidth]{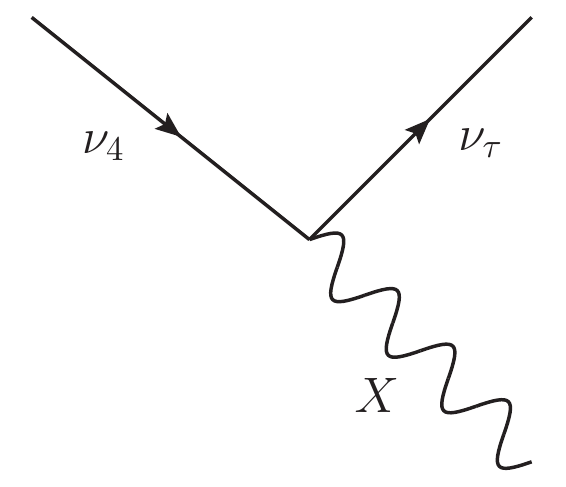}\label{fig_Sterile_Xboson_01}} \,\,
\subfloat[$\nu_4 \to \nu_l e^+ e^-$]
{\includegraphics[width=0.23\textwidth]{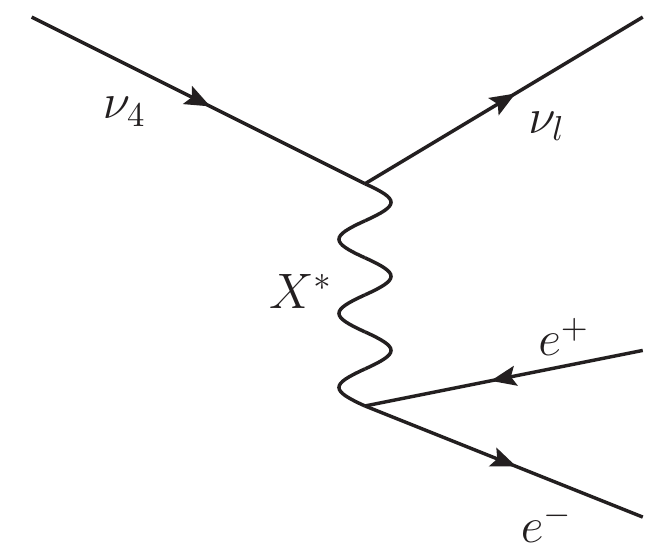}\label{fig_Sterile_Xboson_02}}
  \caption{The  decay channels of the sterile neutrino to virtual and real $X$ boson when $m_4>m_X$ (left) and $m_4<m_X$ (right).
 } 
\end{figure}

\subsection{Width with a new light gauge boson}

When the new $U(1)_X$ interactions are turned on, there are other decay channels to real or virtual $X$ boson as are depicted in Fig.~\ref{fig_Sterile_Xboson_01} and Fig.~\ref{fig_Sterile_Xboson_02} for $m_4>m_X$ and $m_4<m_X$ cases, respectively. 
Depending on the $g_X$ these decay channels can be dominant over the SM CC/NC channels. 
\begin{itemize}
\item {\bf (Model A, $m_4>m_X$):}  As long as $g_X \gsim 10^{-6}$,
the sterile neutrino $\nu_4$ dominantly decays into $X$ boson and active 
neutrino $\nu_\tau$ (Fig.~\ref{fig_Sterile_Xboson_01}).    The decay width is given by
\begin{eqnarray}
\Gamma_{\nu_4 \rightarrow \nu_\tau X}= \frac{m_4 g_X^2 | U_{\tau 4} |^2}{32\pi} \Bigl ( 1 - \frac{m_X^2}{m_4^2} \Bigr ) \Bigl (1 + \frac{m_4^2}{m_X^2} - \frac{2 m_X^2}{m_4^2} \Bigr ). \nonumber \\
\end{eqnarray}
The produced $X$ boson subsequently decays into neutrinos, leptons or neutral vector mesons. 

\item  {\bf (Model B, $m_4<m_X$):} The two-body decay channel is kinematically forbidden and the three-body decay mediated off-shell $X$, $\nu_4 \to \nu_l e^- e^+$ (Fig.~\ref{fig_Sterile_Xboson_02}), is now dominant. The corresponding decay width is 
\begin{eqnarray}
\Gamma_{\nu_4 \rightarrow \nu_\tau e^- e^+}  =  \frac{G_X^2 \epsilon_{\gamma X}^2}{48 \pi^3} | U_{\tau 4}|^2 &&m_4^5 \Bigl [ I_2 (x_{\nu_\tau},x_e,x_e) \nonumber \\
& & +  2 I_{1} (x_{\nu_\tau},x_e,x_e) \Bigr ]
\end{eqnarray}
where $G_X = g_X^2 / (4\sqrt{2} m_X^2)$. Also, there might exist other decay channels, depending on the sterile neutrino mass $m_4$.
\end{itemize}

\begin{figure}[h]
\centering
\subfloat[model A, $m_4>m_X$]
{\includegraphics[width=0.40\textwidth]{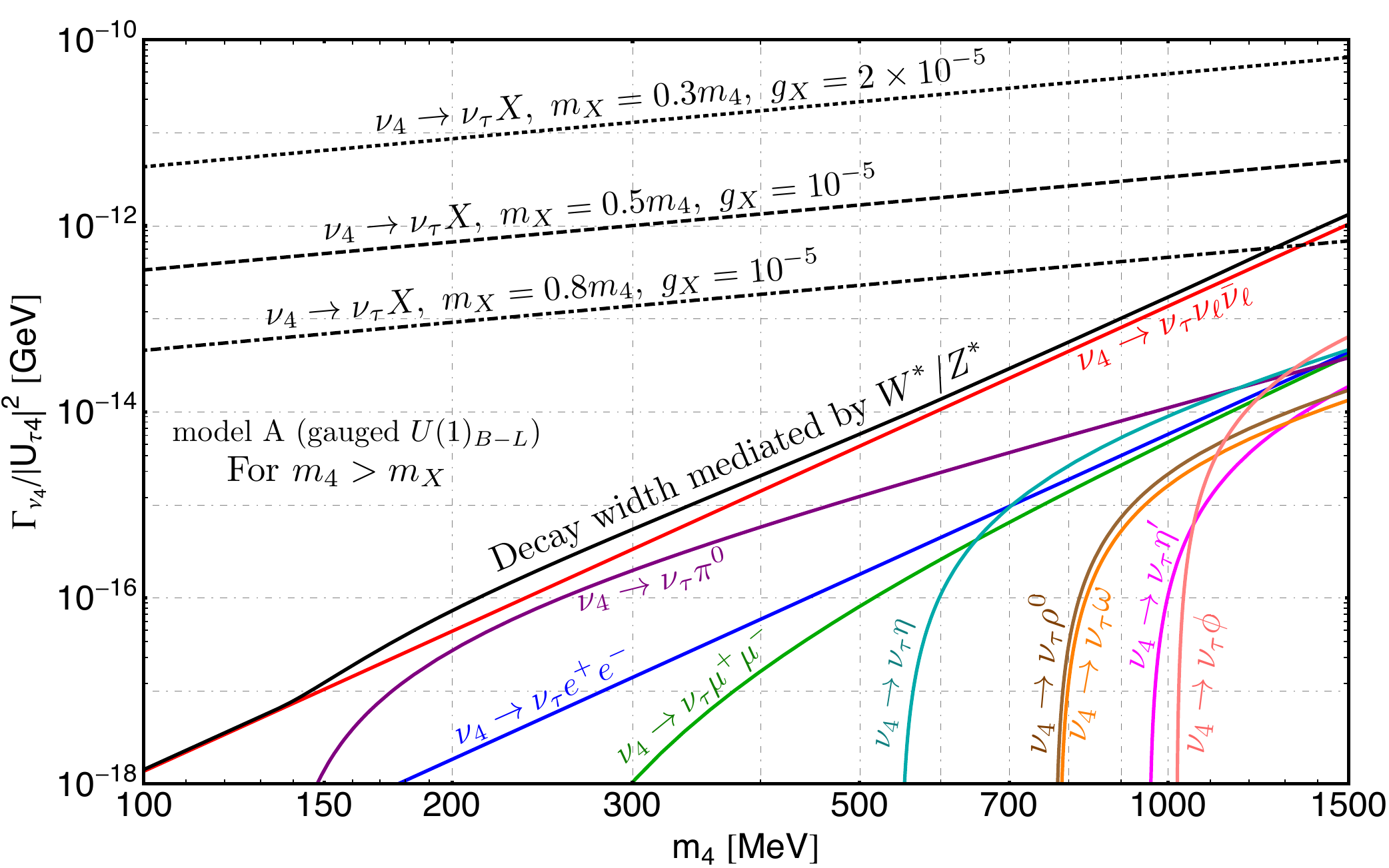}\label{fig_ModelA_nu4_total_width}} \,\,
\subfloat[model B, $m_4<m_X$]
{\includegraphics[width=0.40\textwidth]{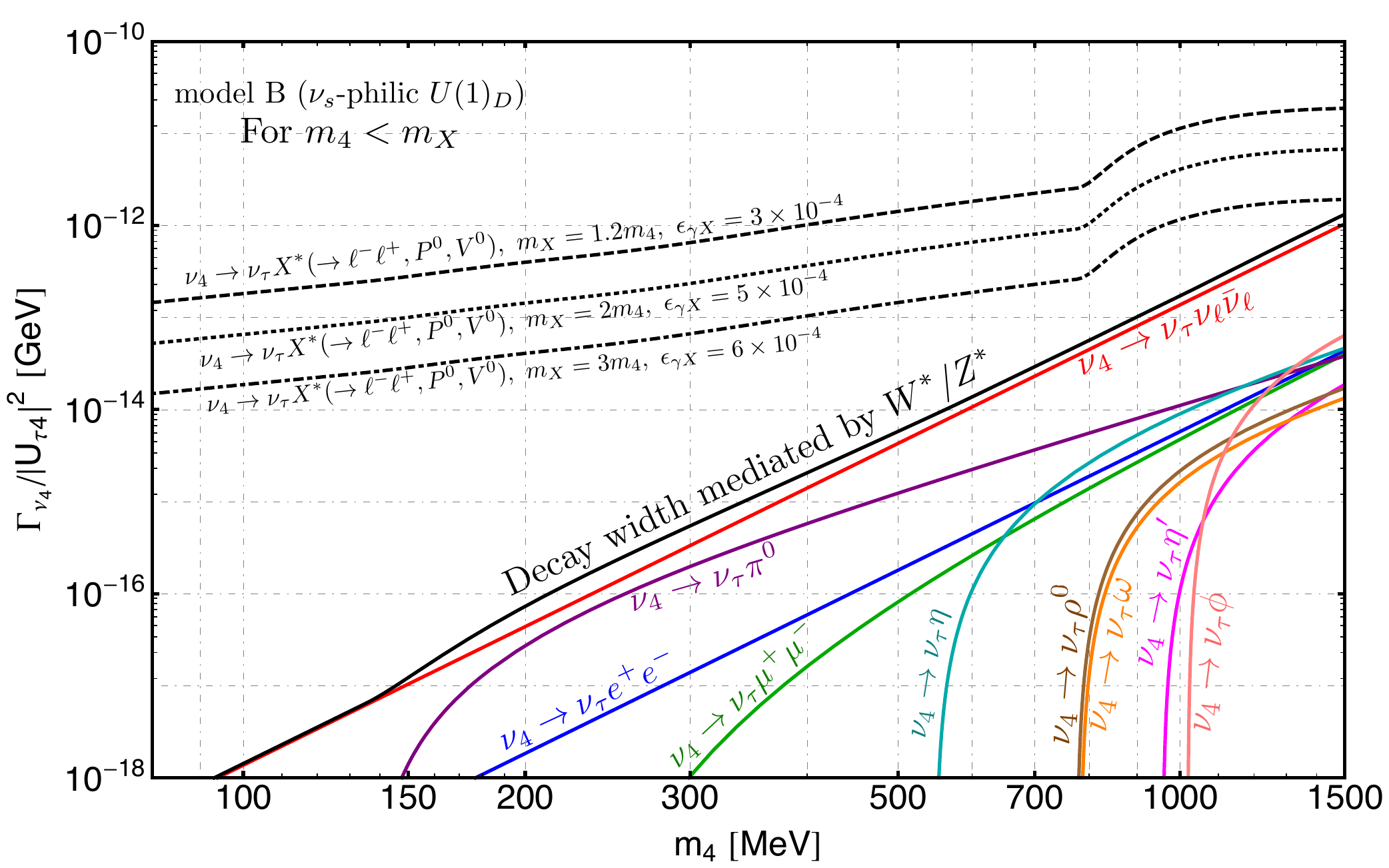}\label{fig_ModelB_nu4_total_width}}
  \caption{The partial width of the sterile neutrino in model A and model B.}\label{fig_nu4_total_width}
\end{figure}

Due to the new gauge interaction, even when the effective coupling constants are small as $g_X \sim 10^{-4} - 10^{-6}$ in model A and $g_X \epsilon_{\gamma X} \sim 10^{-4} - 10^{-3}$ in model B, the total width of sterile neutrino is $\sim 10^2 - 10^4$ times larger than the conventional width by SM NC/CC interactions as one can see in Fig.~\ref{fig_nu4_total_width}. The solid lines correspond to the SM contributions. The dotted lines are related to new decay channel in the presence of the light $X$ boson.

\section{Constraints and expected sensitivities on $|U_{\tau4}|^2$}
\label{sec:constraints}
With the larger decay width of $\nu_4$ with the new gauge interactions, the allowed parameter space of $U_{\ell 4}$ should be reconsidered. We focus on $U_{\tau 4}$ here to be definite but our method should be applicable to other mixing parameters as well.

\subsection{The long-lived particle searches}

The expected number of signal events $N_{\rm sig}$ for the fiducial decay region of experiment $[L-\Delta, L]$ is
\begin{eqnarray}
N_{\rm sig.} & = & \int_{E_{\min}}^{E_{\max}} dE_{\nu_4} \left [ \frac{dN_{\nu_4}(E_{\nu_4})}{dE_{\nu_4}} \times \left ( e^{- \frac{L-\Delta}{d} } - e^{- \frac{L}{d} } \right ) \ \ \  \right . \nonumber \\
& & \left . \ \ \ \ \ \ \ \ \times \text{Br}(X /\nu_4 \to \text{visible}) \times A_{\rm eff} (E_{\nu_4}) \frac{}{} \right ], \ \ \ \ \ \ 
\end{eqnarray}
where $dN_{\nu_4}/dE_{\nu_4}$ is the (model dependent) flux of sterile neutrinos entering into fiducial decay region and $d$ is the decay width of $\nu_4$. More explicitly, $d = \gamma_{\nu_4} c \tau_{\nu_4} + \gamma_X c \tau_X$ when $m_4 > m_X$, and $d = \gamma_{\nu_4} c \tau_{\nu_4}$ when $m_4 < m_X$, where $\gamma_\alpha$ is the Lorentz factor for the corresponding particle $\alpha=\nu_4, X$. The detection efficiencies $A_{\rm eff}(E_{\nu_4})$ for each experiment are given in the references in Table.~\ref{Table_LongLivedParticleSearches}.
%
The sterile neutrinos are produced via the mixing with the active neutrino so that its flux is  given as 
\begin{eqnarray}
\frac{dN_{\nu_4}}{dE_{\nu_4}} & \approx & \frac{dN_{\nu_\tau}}{dE_{\nu_\tau}}  \times | U_{\tau 4} |^2 \times (\text{phase space suppresion}) \nonumber \\
& & \times (\text{helicity suppression}).
\end{eqnarray}
where the phase space and helicity suppression factors are from the nonzero sterile neutrino masses~\cite{Orloff:2002de}. 
Here we assume $U_{\tau 4} \gg {\rm Max}[ U_{\mu 4}, U_{e4}]$ so that only tau neutrinos are taken into account. Tau neutrinos are copiously produced from the decay of mesons, as $D_s \to \tau \nu_\tau,~ B\to D \tau \nu_\tau$. 

\begin{table*}[t]
\centering
\begin{tabular}{| c | c | c | c | c | c | c | c | c | c |}
\hline
\ \ \ \ \ \ \ Experiment \ \ \ \ \ \ \ & \ \ $N_{\rm POT}$ or $\int \mathcal{L} \ dt$ \ \ & \ \ $\sqrt{s}$ \ \ & \ $E_{p \text{ beam}}$ \ & \ \ $L$ \ \ & \ \ $\Delta$ \ \ & \ $\langle E_{\nu_4} \rangle$ \ & \ 95\% C.L. limit \ & \ \ Ref. \ \ \\
\hline
\hline
FASER (LHC Run 3)& $\int \mathcal{L} \ dt = $ $150$ fb$^{-1}$ & \multirow{2}{*}{\ 14 TeV \ } & \multirow{2}{*}{\ \ \ 7 TeV \ \ } &  \ \multirow{2}{*}{480m} \ & \ \multirow{2}{*}{1.5m} \ & \ \multirow{2}{*}{$ \sim 1 \text{ TeV} $} \ & \ \ \multirow{5}{*}{ $N_{\rm sig} \geq 3$ } \ \ &\multirow{2}{*}{ \cite{Ariga:2018uku, Abreu:2019yak, Kling:2018wct} } \\
\cline{1-2}
FASER2 (HL-LHC) & $\int \mathcal{L} \ dt = $ $3$ ab$^{-1}$ & & & & & & & \\
\cline{1-7} \cline{9-9}
SHiP & $N_{\rm POT} = 2 \times 10^{20}$ & \multirow{2}{*}{27.4 GeV} & \multirow{2}{*}{400 GeV} & 110m & 50m & & & \cite{Alekhin:2015byh, Bonivento:2013jag} \\
\cline{1-2} \cline{5-6} \cline{9-9}
CHARM &$N_{\rm POT} = 2.4 \times 10^{18}$ & & & 515m & 35m & \ $\sim 50$ GeV \ & & \cite{Orloff:2002de, GonzalezGarcia:1996ri} \\
\cline{1-6} \cline{9-9}
NOMAD & $N_{\rm POT} = 4.1 \times 10^{19}$ & 29 GeV & 450 GeV & 835m & 290m & & & \cite{Astier:2001ck} \\
\hline
\end{tabular}
    \caption{The fiducial region, collision energies and the number of protons on target (POT) of long-lived particle searches in fixed target/head-on collision experiment. For the details on sterile neutrino spectrum, see the references on the ``Ref." column.}
    \label{Table_LongLivedParticleSearches}   
\end{table*}

\subsubsection{CHARM/NOMAD}

From the beam-dump searches using proton beams from CERN SPS, CHARM (using $400$ GeV$/c$ proton beam and copper target) \cite{Orloff:2002de} and NOMAD (using $450$ GeV$/c$ and beryllium target)  \cite{Astier:2001ck} have provided the exclusion limits on active-sterile mixing for tau flavor $| U_{\tau 4} |^2$ in the range of sterile neutrino mass $m_4 = 10-290$ MeV (CHARM) and $m_4 = 10-190$ MeV (NOMAD), using the signal $\nu_4 \to \nu_\tau e^- e^+$. Without $X$ boson effects,  the largish parameter region up to $|U_{\tau4}|^2 \gsim 10^{-4}$ were probed.\cite{Orloff:2002de, Astier:2001ck} but with the $X$ boson, even smaller mixing $| U_{\tau 4} |^2 \sim 10^{-4} - 10^{-8}$ can be reached by the experiments.  The recasted results of Model A and Model B are shown in   Fig.~\ref{fig_BP_modelA} and  Fig.~\ref{fig_BP_modelB}, respectively, for the same mass ranges of sterile neutrino.

\subsubsection{FASER}

FASER \cite{Ariga:2018uku} (and its upgrade FASER2) is designed to search for long-lived particles (LLP). The LLPs are produced in the forward direction of the proton-proton head-on collision with $\sqrt{s} = 14$ TeV at the interaction point (IP) of ATLAS detector then reach to the FASER detector with $L \sim$ 480m and $\Delta \sim$ 1.5m.  FASER (FASER2), planned to use 150 fb$^{-1}$ (3 ab$^{-1}$) data of LHC Run 3 (HL-LHC), has good sensitivity to active-sterile neutrino mixings $| U_{\tau 4} |^2$, due to its large number of produced neutrinos \cite{Abreu:2019yak}. 
We show the FASER/FASER2 limits of our models in Fig.~\ref{fig_BP_modelA} and Fig.~\ref{fig_BP_modelB}. Compared to the SHiP experiment, which has advantage in the small mixing region $| U_{\tau 4} |^2 \sim 10^{-4} - 10^{-8}$, FASER have strength in  heavier mass region $300 \text{ MeV} \lsim m_4 \lsim 1.67$ GeV.

\subsubsection{SHiP}

SHiP \cite{Alekhin:2015byh} is proposed to search for LLP using CERN SPS proton beam of 400 GeV and the molybdenum target.
SHiP has strength in proton on target (POT) about $2 \times 10^{20}$, which is larger than the older searches such as CHARM and NOMAD
and also the expected background is claimed to be significantly smaller~\cite{Bonivento:2013jag}. We indicate the expected limits from SHiP search for Model A in Fig.~\ref{fig_BP_modelA} and for Model B in Fig.~\ref{fig_BP_modelB}. Thanks to the large POT, the sensitivity can be low down to the seesaw expected active-sterile mixing values $| U_{\tau 4} |^2 \sim 10^{-8} - 10^{-10}$.

\subsection{Collider searches at $Z$ decays}

Regardless of its flavor $\ell$, the active-sterile mixings $U_{\ell 4}$ can be probed by $Z$-pole search at LEP-I experiment, including the monojet search of $Z$ decays ($Z \to j + \slashed{E}_T$) and invisible $Z$ decay width measurement.

\subsubsection{LEP-I monojet search}

DELPHI reported the weak isosinglet neutral heavy lepton ($\nu_4$) search with $3.3\times 10^6$ $Z$ bosons at LEP-I experiment. 
Several separate searches have been performed e.g., for promptly decaying $\nu_4 \to \text{monojet}$ ($m_4 \gsim 6$ GeV), and for long-lived $\nu_4$ giving secondary vertices ($1 \text{ GeV} \lsim m_4 \lsim 6$ GeV)~ \cite{Abreu:1996pa} and provided the bound on the branching fraction of $Z$ boson as
\begin{eqnarray}
\text{Br}(Z \to \nu \nu_4) & < & 1.3 \times 10^{-6}\,\,(\text{$95\%$ C.L.)}
\end{eqnarray}
 in a wide window $m_4 = 3.5 - 50$ GeV. Using the relation $\text{Br}(Z \to \nu \nu_4) =\text{Br}(Z \to \nu_\ell \bar{\nu}_\ell) |U_{\ell 4}|^2 \left(1-\frac{m_4^2}{m_Z^2}\right)^2 \times \left(1+\frac{m_4^2}{2m_Z^2}\right)$ and $\text{Br}(Z \to \nu_\ell \bar{\nu}_\ell)\approx 0.063$, we read the bound on the active-sterile mixing, $| U_{\ell 4} |^2 \lsim 2.1 \times 10^{-5}$ when $m_4 \ll m_Z$. 

In Model A, $m_4 > m_X$, $\nu_4$ is short lived. The branching fraction of $X$ boson also changes the interpretation.  The limit 
is now read as depicted in Fig.~\ref{fig_BP_modelA}:
\begin{eqnarray}
| U_{\ell 4} |^2 & < & \frac{2.1 \times 10^{-5}}{\text{Br}(X \to \text{hadrons})} .
\end{eqnarray}

In Model B, $1~{\rm GeV} < m_4 < m_X$, the $X$ boson mediated channels ($\nu_4 \to \nu_\ell X^*, \ X^* \to e^- e^+, \mu^- \mu^+ , \text{hadrons}$) are subdominant. Therefore the bound is similar to the original DELPHI result as can be seen in Fig.~\ref{fig_BP_modelB}. 

\subsubsection{$Z$ boson invisible width}

The upper limit of active-sterile mixing $U_{\ell 4}$ ($\ell=e,\mu,\tau$) are given by 
\begin{eqnarray}
| U_{l 4} |^2 & < & \frac{1}{\text{Br}(X \to \text{invisible})}  \cdot \left ( \frac{\Gamma_{Z \to \text{invisible}}^{\rm Exp.}}{\Gamma_{Z \to \text{invisible}}^{\rm SM}} - 1 \right ), \ \ \ \ \ \ 
\end{eqnarray}
where experimental observation of invisible $Z$ width at LEP and its SM prediction are \cite{ALEPH:2005ab,Voutsinas:2019hwu,Janot:2019oyi}
\begin{eqnarray}
\Gamma_{Z \to \text{invisible}}^{\rm Exp.} & = & 499.0 \pm 1.5 \text{ MeV}, \\
\Gamma_{Z \to \text{invisible}}^{\rm SM} & = & 501.69 \pm 0.06 \text{ MeV}.
\end{eqnarray}

For model A (gauged $U(1)_{B-L}$), we get
\begin{eqnarray}
|U_{\ell 4}|^2 & \lsim & 0.0072
\end{eqnarray}
as the 3$\sigma$ upper limit, fixing the branching ratio of $X$ boson as $\text{Br}(X \to \text{invisible}) \simeq 50 \%$. 

For model B (sterile $\nu$-philic $U(1)_s$), the invisible decay branching ratio of sterile neutrino $\text{Br}(\nu_4 \to \nu_\ell \nu_{\ell'} \bar{\nu}_{\ell'})$ suppressed by $| U_{\ell 4} |^4$ compared to visible channel branching ratios $\sim \mathcal{O}(1)$, and the bound becomes
\begin{eqnarray}
| U_{l 4} |^6 & \lsim & \left ( \frac{\Gamma_{Z \to \text{invisible}}^{\rm Exp.}}{\Gamma_{Z \to \text{invisible}}^{\rm SM}} - 1 \right ), \ \ \ \ \ \ 
\end{eqnarray}
or $| U_{\ell 4} |^2 < 0.153$ for $m_4 \lsim 1$ GeV. For heavier masses ($m_4 \gsim 1$ GeV), $\text{Br}(\nu_4 \to \nu_\ell \nu_{\ell'} \bar{\nu}_{\ell'}) \approx 1$ and the bound becomes $| U_{\ell 4} |^2 < 0.0036$.

\subsection{Collider search at B-factories}

\subsubsection{B-factories ($\tau$ decays)}

Due to its clear environment for the reconstruction, $\tau^- \to \nu_4 \pi^- \pi^+ \pi^-$ is the most promising channel to observe the sterile neutrino from tau decays in B-factory experiment such as Belle and BaBar. We adopt a bound from Ref.~\cite{Kobach:2014hea} and rescaled it with invisible branching fraction of $X$ boson (in model A) or $\nu_4$ (in model B).

\subsubsection{B-factories ($B$ decays)}

As well as $\tau$ decays, the decay of $B$ mesons $B \to D\tau \nu_4$ can lead a limit of active-sterile mixing $| U_{\tau 4} |^2$. As pointed out in \cite{Kim:2019xqj}, there is a potential source of background $B \to D^* \tau \nu_4, D^* \to D \gamma_{\rm miss}$ for sterile neutrino lighter than 1 GeV. Thus, we adopt the limit from Ref.~\cite{Kim:2019xqj} only for $m_4 \gsim 1$ GeV.

\subsection{Neutrino Telescopes}

IceCube neutrino telescope also probes sterile neutrinos.  The search of double-cascade events with atmoshperic ($E_\nu \sim 5-50$ GeV) and astrophysical ($E_\nu \sim 60 \text{ TeV} - 1$ PeV) neutrinos are our targets. 

\subsubsection{IceCube-DeepCore double cascade search (atmospheric $\nu$)}

As precisely measured by e.g. Super-Kamiokande \cite{Richard:2015aua}, the muon neutrinos are copiously produced in atmosphere. The muon neutrino can convert to the our target (tau neutrinos) by oscillation with the probability:
\begin{eqnarray}
P(\nu_\mu \to \nu_\tau) & = & \sum_{j,k} U_{\mu j} U_{\tau j}^* U_{\mu k}^* U_{\tau k} \exp \left (i \frac{\Delta m_{jk}^2 L}{2 E_\nu} \right ) \ \ \ \ \ \nonumber \\
& \approx & \cos^4 \theta_{13} \sin^2 \theta_{23} \sin^2 \left ( \frac{\Delta m_{jk}^2 L}{4 E_\nu} \right ), 
\end{eqnarray}
where $U$ are the elements of the $3 \times 3$ PMNS matrix. One can notice that the conversion is maximized when $L= 2 R_\oplus$, $E_\nu \sim 25$ GeV.

IceCube-DeepCore (inside the IceCube detector volume) has been designed to detect neutrinos with $E_\nu = 1-100$ GeV and the effective mass $\sim 10-30$ Mton, at $2100-2400$m underground \cite{Aartsen:2016psd}. 
From the 2015-2016 data, about $N_{\rm NC}^{\nu_\tau} = 1.4 \times 10^4$ NC tau neutrino events ($\nu_\tau N \to \nu_\tau N'$) in $E_\nu = 5.6 - 56$ GeV \cite{Aartsen:2019tjl} have been analyzed. Due to the $U_{\tau 4}$ mixing, there are $\nu_\tau N \to \nu_4 N'$ events but with the suppressed rate by $|U_{\tau4}|^2$. Once produced, $\nu_4$ will fly about $20$ m then leave a a unique event topology of double-cascade. The event number is estimated as
\begin{eqnarray}
N_{\rm sig.} & \simeq & \int dE_\nu \Bigl [ \frac{dN_{\rm NC}^{\nu_\tau}}{dE_\nu}  \times \left ( e^{- \frac{L-\Delta}{d} } - e^{- \frac{L}{d} } \right ) \nonumber \\
& & \times \text{Br}(X/\nu_4 \to \text{visible}) \Bigr ],
\end{eqnarray}
where $L=300$m is the fiducial vertical length of the DeepCore and $L-\Delta = 20$m is the minimum length to distinguish the double-cascade event from the background events.  A similar analysis without the $X$ boson was done in Ref. \cite{Coloma:2017ppo}. 
We request $N_{\rm sig.} \geq 10$ to set the limit.

\subsubsection{IC-Gen2 double cascade search (astrophysical $\nu$)}

The sterile neutrinos are produced from the astrophysical diffuse tau neutrinos. From the 7.5 years data of  IceCube for high-energy starting events (HESE), the flux of the astrophysical tau neutrino is obtained as \cite{Williams:2018kpe}
\begin{align}
\frac{d\Phi_{\nu + \bar{\nu}}}{dE_\nu} & = \Phi_0 \times 10^{-18} \left ( \frac{E_\nu}{100 \text{ TeV}} \right )^{-\gamma} \nonumber \\
\ &\left [ \text{GeV}^{-1} \text{cm}^{-2} \text{s}^{-1} \text{sr}^{-1} \right ], 
\end{align}
where $\Phi_0 = 6.45_{-0.46}^{+1.46}$ and $\gamma = 2.89_{-0.19}^{+0.20}$, for the sum of all flavors of light SM neutrinos ($\ell = e,\mu,\tau$). The equal flavor composition ($\nu_e : \nu_\mu : \nu_\tau = 1:1:1$) is assumed for astrophysical neutrinos. 
The astrophysical neutrinos can interact with nucleons in the IceCube detector and the sterile neutrinos are produced by the process $\nu_\tau N \to \nu_4 N'$. The sterile neutrino decays by $\nu_4 \to \nu_\tau X/ \nu_\tau X^*$  as discussed in the previous subsection. 
In the range $E_\nu = 60 \text{ TeV} - 1$ PeV, the main candidate of the SM background is the tau-induced double cascade \cite{Stachurska:2019wfb}, which has the characteristic ratio between the deposited energy $E_{\rm dep.}$ and the separation of two cascades $L_{\rm dec.}$ as $E_{\rm dep.}/{L_{\rm dec.}} |_{\tau\text{-induced}} \sim 1 \text{PeV}/{50 \text{m}}$. On the other hand, the ratio is $E_{\rm dep.}/{L_{\rm dec.}} |_{\nu_4\text{-induced}} \sim 1 \text{TeV}/{\text{m}}$ for the signal events. Therefore, we conclude that the signal events are distinguishable  from the background events. Unfortunately, however, the expected signal events are limited by the flux of the astrophysical neutrinos thus we only get a weaker bound compared to other constraints.

\begin{figure*}[ht]
\centering
\subfloat[$(m_X, \ g_X) = (0.1 m_4, \ 10^{-5})$]
{\includegraphics[width=0.32\textwidth]{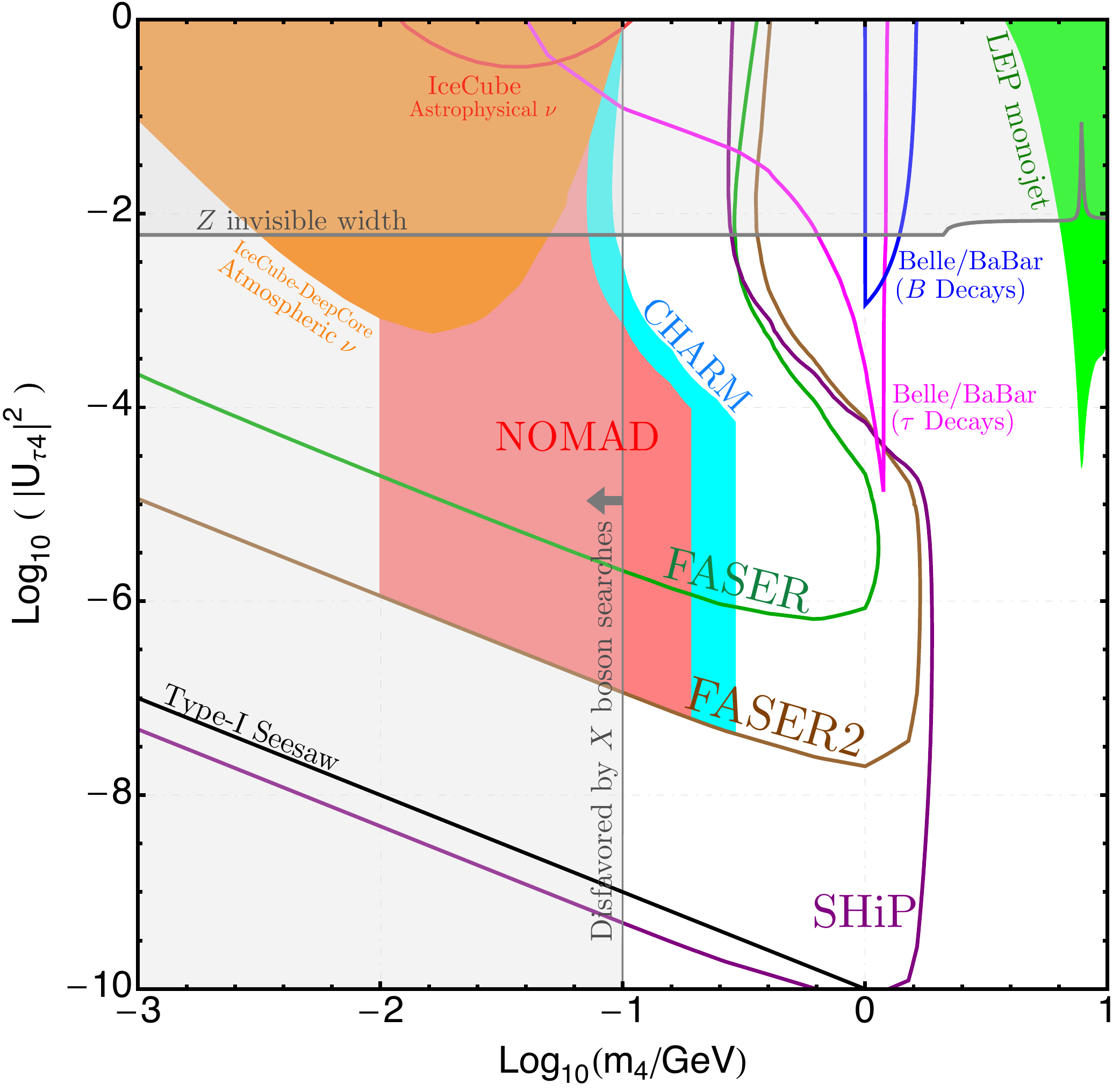}\label{fig_BP1_modelA}} \,\,
\subfloat[$(m_X, \ g_X) = (0.3 m_4, \ 10^{-5})$]
{\includegraphics[width=0.32\textwidth]{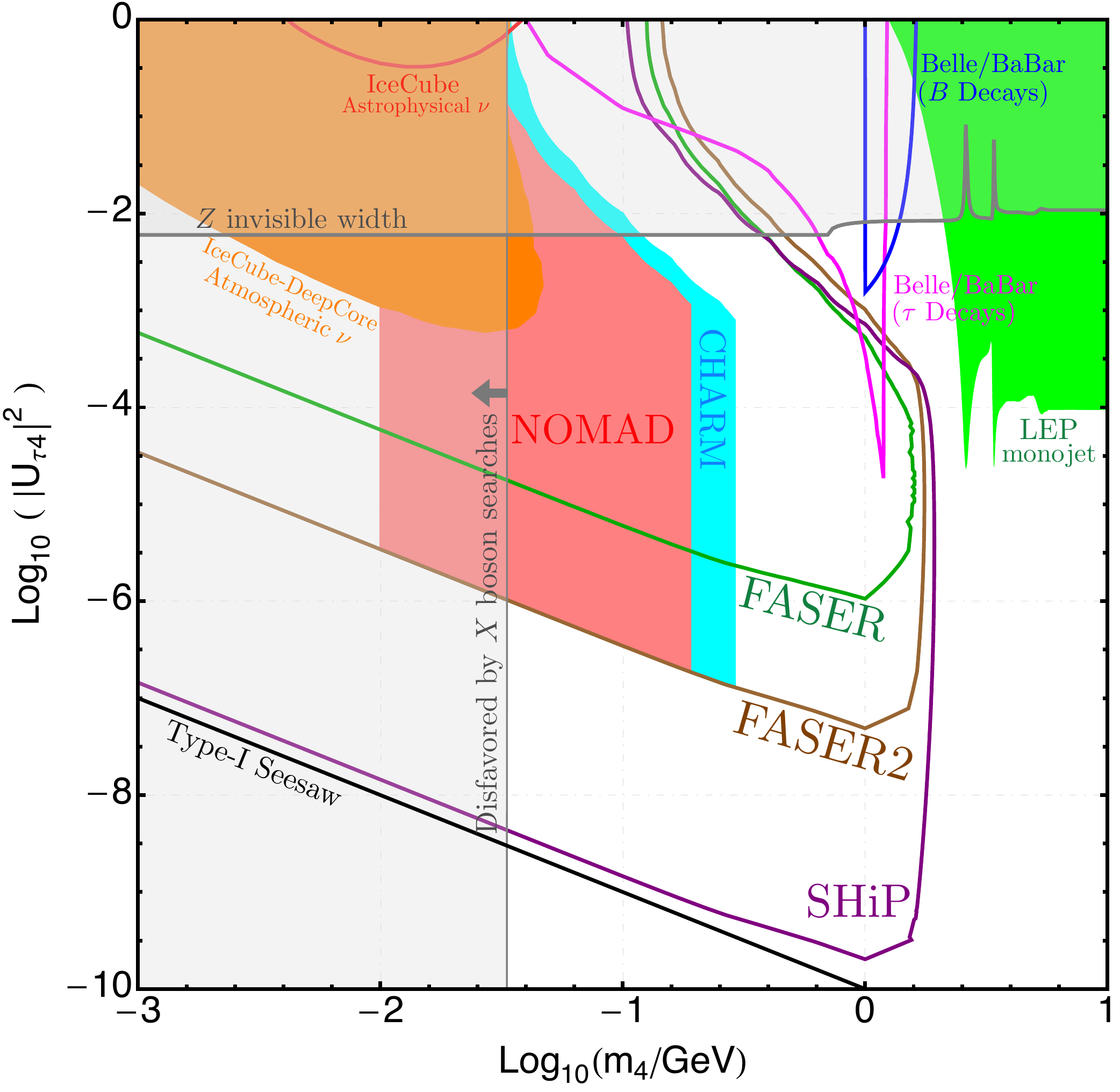}\label{fig_BP2_modelA}} \,\,
\subfloat[$(m_X, \ g_X) = (0.5 m_4, \ 10^{-5})$]
{\includegraphics[width=0.32\textwidth]{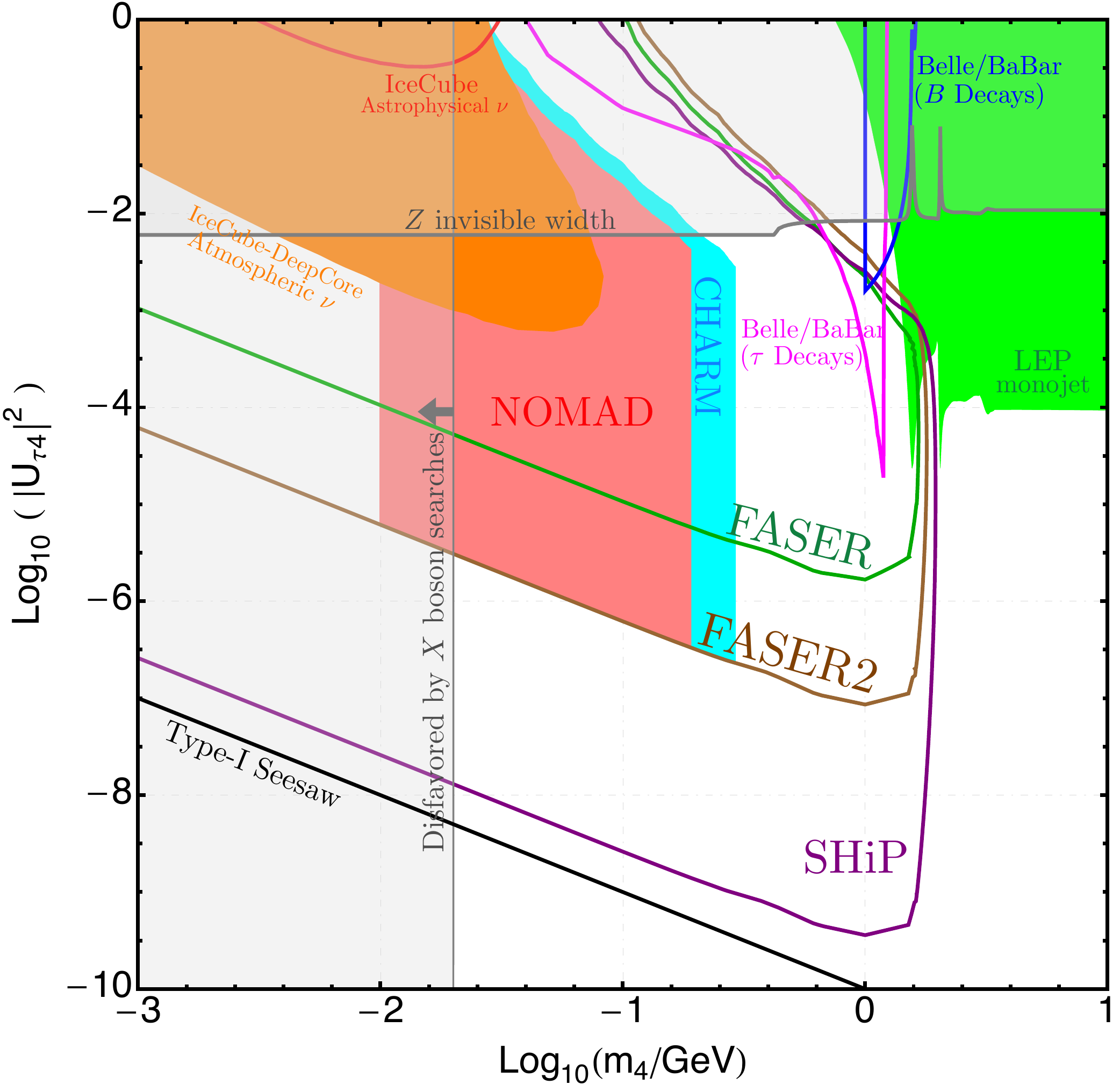}\label{fig_BP3_modelA}} \,\,
\subfloat[$(m_X, \ g_X) = (0.8 m_4, \ 10^{-5})$]
{\includegraphics[width=0.32\textwidth]{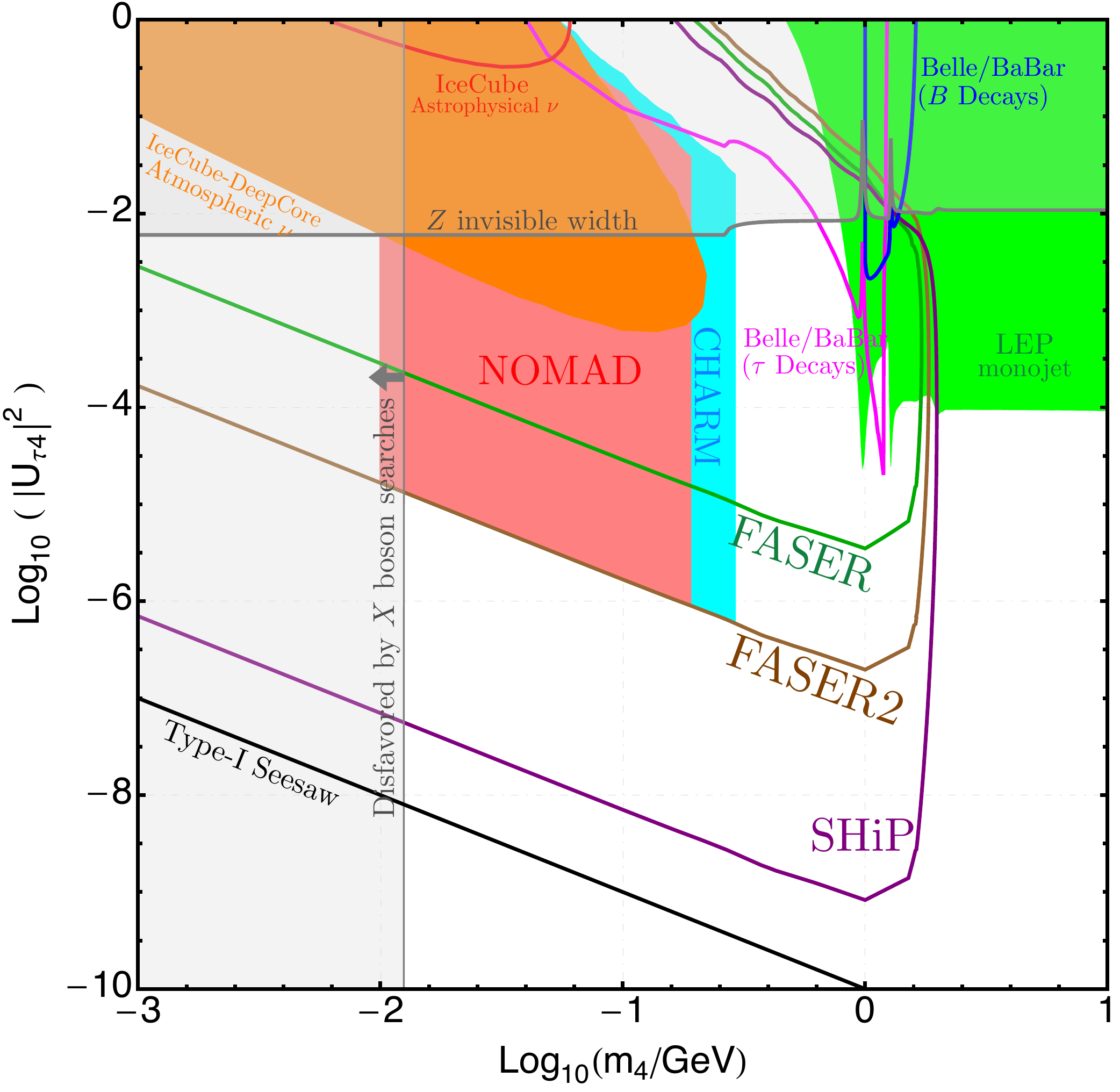}\label{fig_BP4_modelA}} \,\,
\subfloat[$(m_X, \ g_X) = (0.3 m_4, \ 10^{-4})$]
{\includegraphics[width=0.32\textwidth]{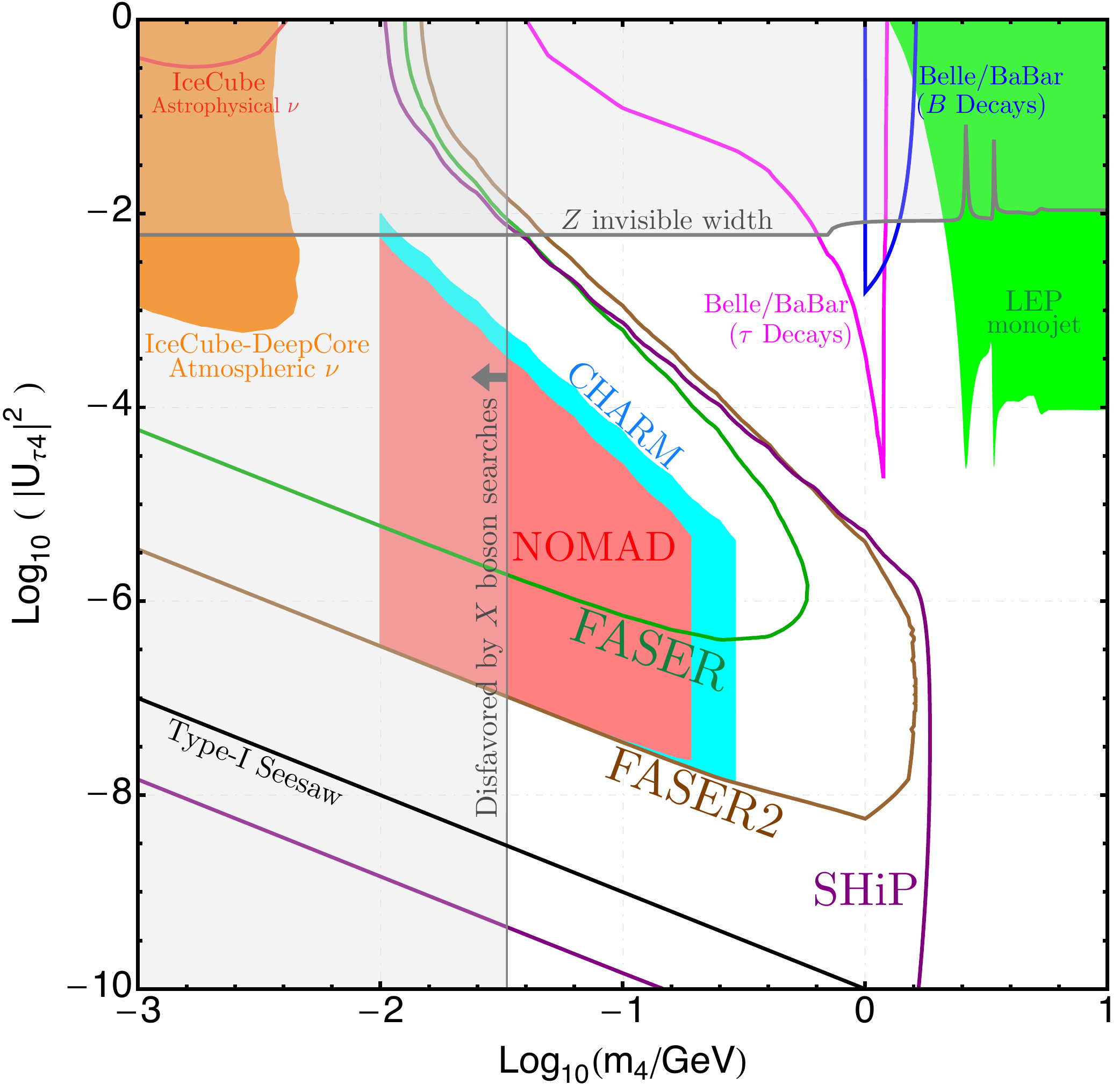}\label{fig_BP5_modelA}} \,\,
\subfloat[$(m_X, \ g_X) = (0.8 m_4, \ 3 \times 10^{-4})$]
{\includegraphics[width=0.32\textwidth]{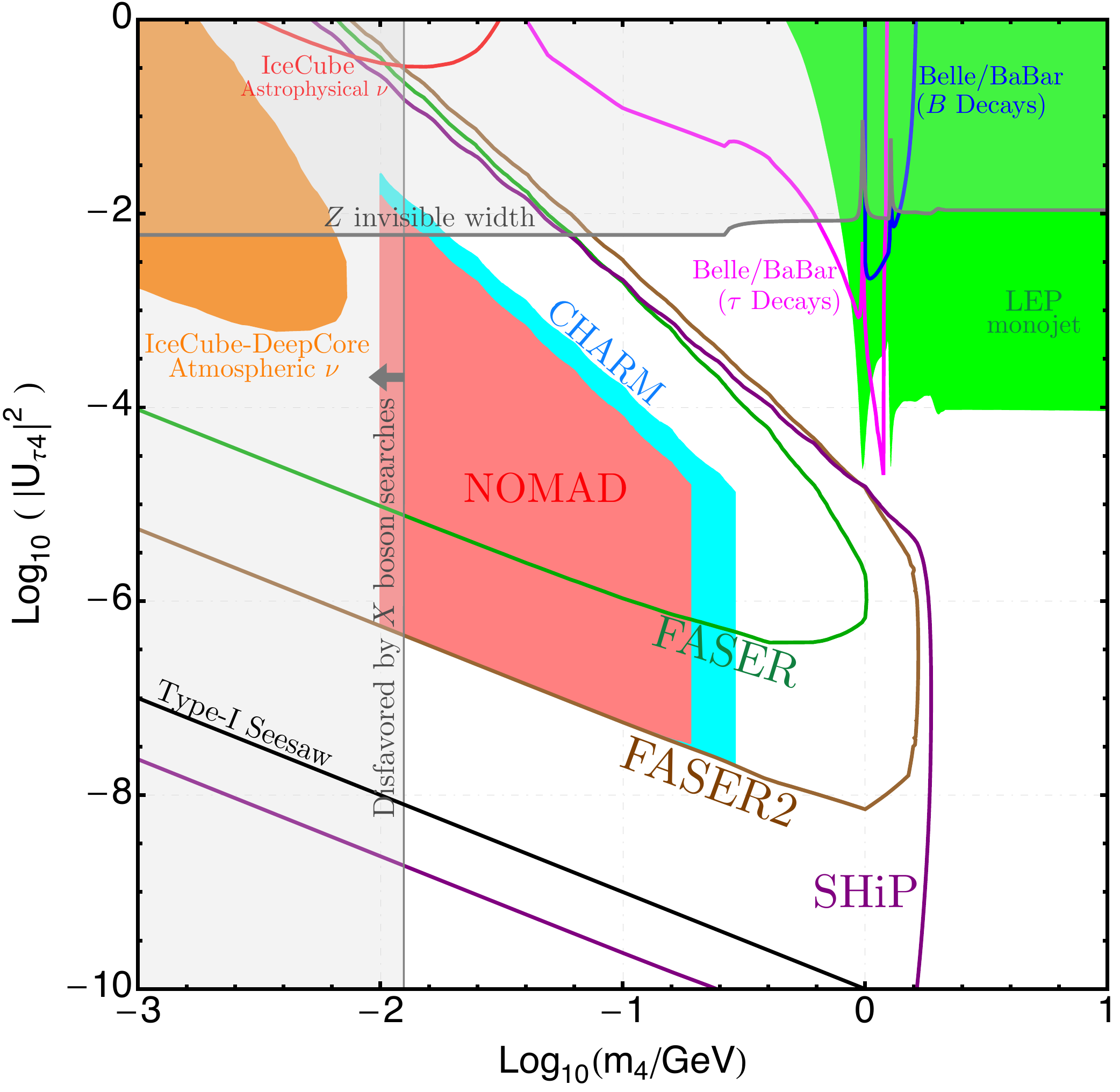}\label{fig_BP6_modelA}} \,\,
  \caption{(Model A) Current constraints on the (tau) active-sterile mixing $| U_{\tau4}|^2$, with varying $B-L$ gauge boson parameters ($m_X, g_X$). The constraints from CHARM (red shaded), NOMAD (cyan shaded), IceCube-DeepCore (orange shaded), LEP monojet (green shaded) and $Z$ boson invisible width (gray shaded with horizontal gray line) are shown. Also, the future expected limits from FASER (green solid line), FASER2 (brown solid line), SHiP (purple solid line), $B\to D \tau \nu_4$ (blue solid line)/$\tau^- \to \nu_4 \pi^- \pi^+ \pi^-$ (magenta solid line) at B-factories and IceCube-Gen2 double cascade search with astrophysical neutrino (red solid line) are shown. Black solid line is the type-I seesaw expected mixing $| U_{\tau4} |_{\rm seesaw}^2 = (0.1 \text{ eV})/m_4$. Vertical gray solid line comes from the lower bound of $m_X$ from experimental searches.}\label{fig_BP_modelA}
\end{figure*}

\begin{figure*}[ht]
\centering
\subfloat[$(m_X, \ \epsilon_{\gamma X}) = (1.2 m_4, \ 3 \times 10^{-4})$]
{\includegraphics[width=0.32\textwidth]{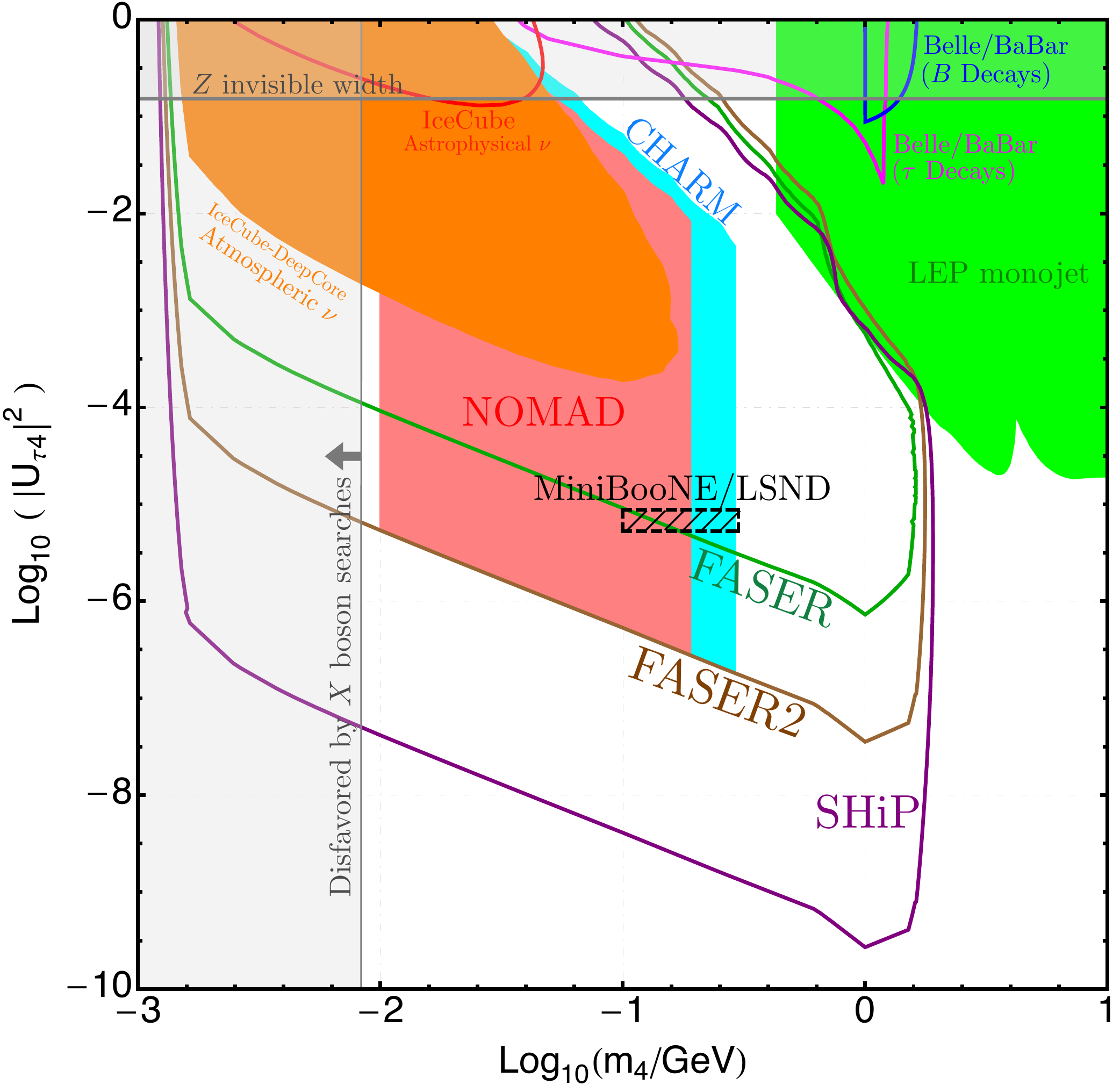}\label{fig_BP1_modelB}} \,\,
\subfloat[$(m_X, \ \epsilon_{\gamma X}) = (2 m_4, \ 5 \times 10^{-4})$]
{\includegraphics[width=0.32\textwidth]{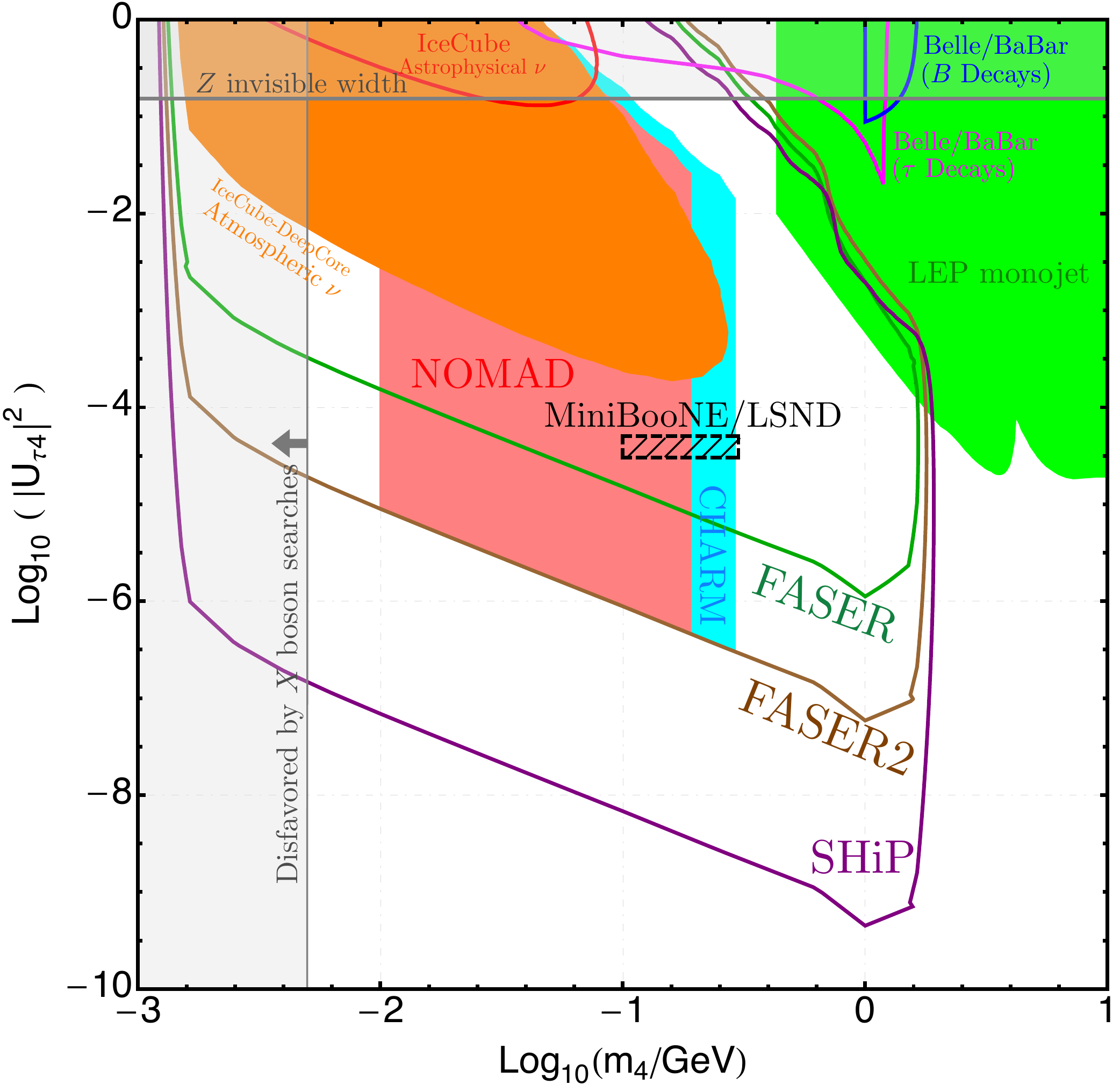}\label{fig_BP2_modelB}} \,\,
\subfloat[$(m_X, \ \epsilon_{\gamma X}) = (3 m_4, \ 6 \times 10^{-4})$]
{\includegraphics[width=0.32\textwidth]{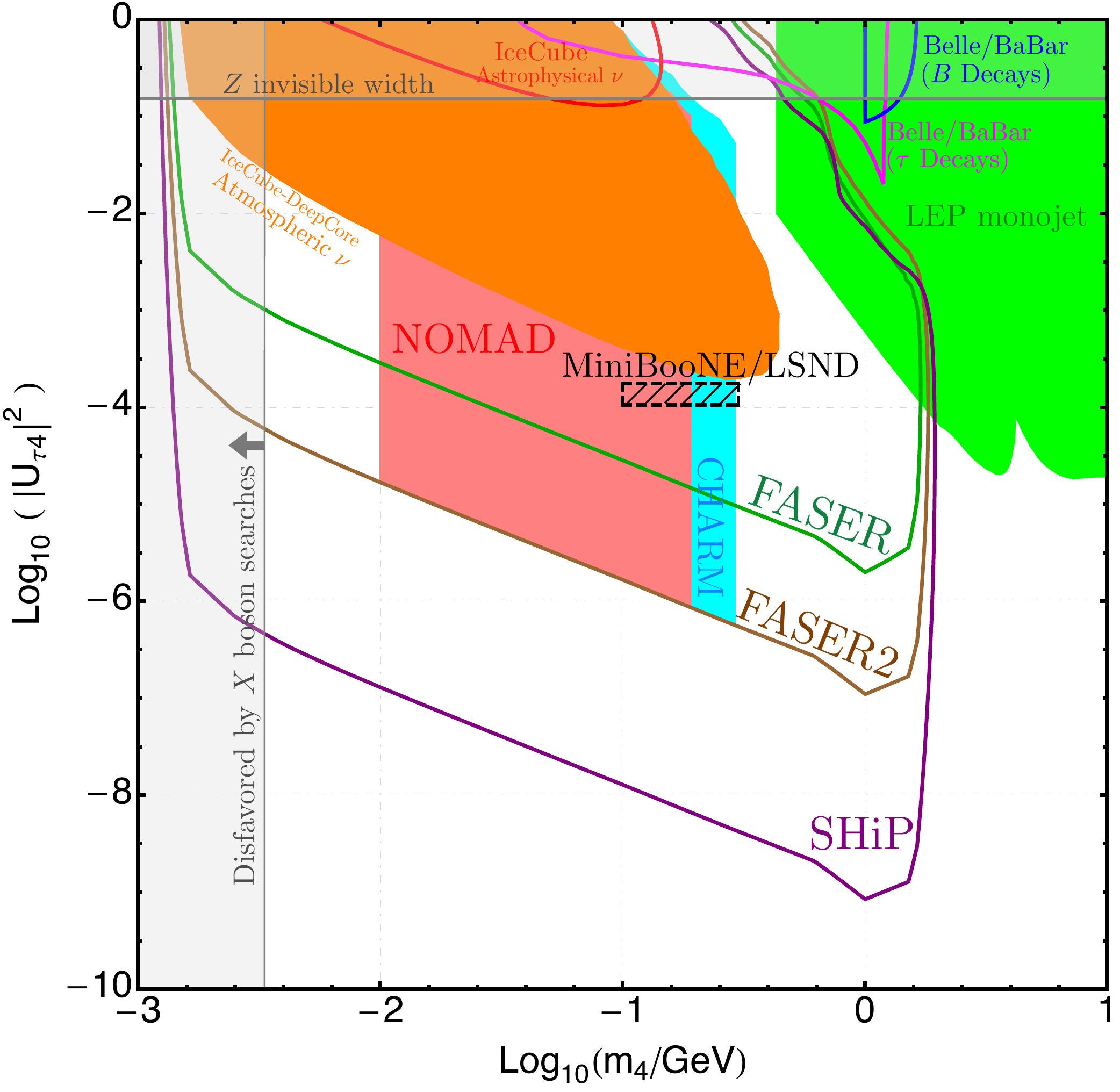}\label{fig_BP3_modelB}} \,\,
\subfloat[$(m_X, \ \epsilon_{\gamma X}) = (5 m_4, \ 8 \times 10^{-4})$]
{\includegraphics[width=0.32\textwidth]{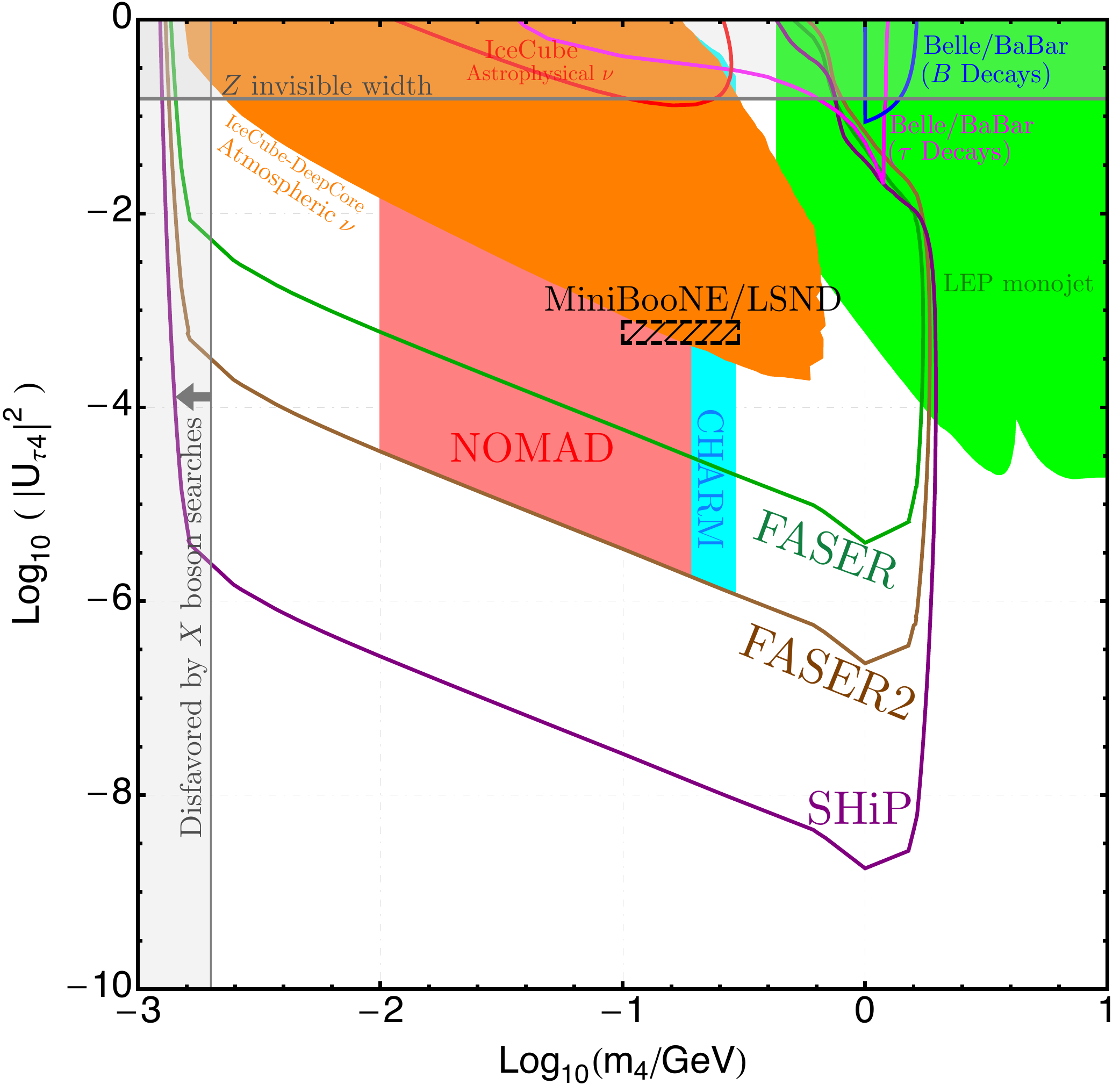}\label{fig_BP4_modelB}} \,\,
\subfloat[$(m_X, \ \epsilon_{\gamma X}) = (8 m_4, \ 1 \times 10^{-3})$]
{\includegraphics[width=0.32\textwidth]{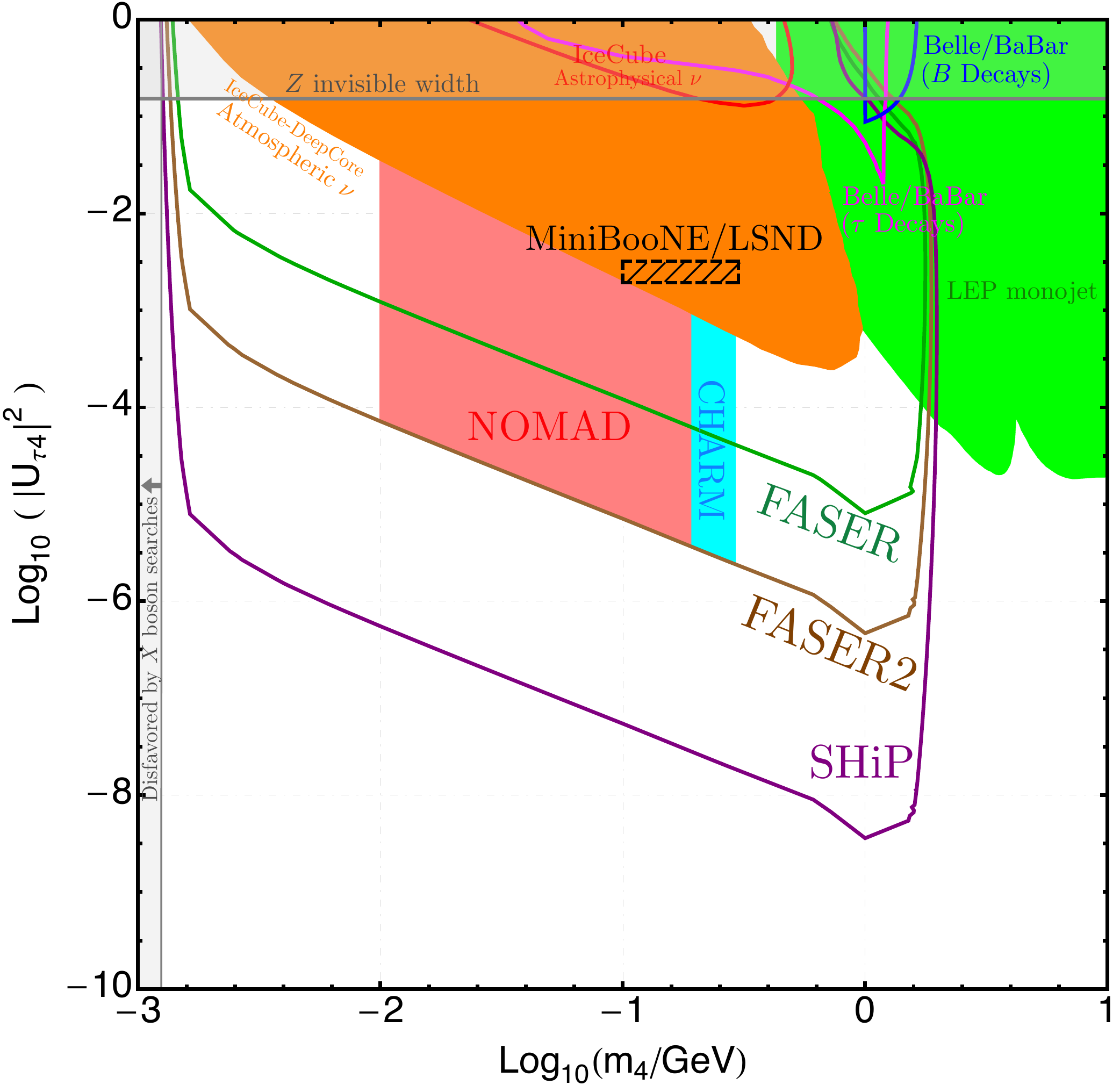}\label{fig_BP5_modelB}} \,\,
\subfloat[$(m_X, \ \epsilon_{\gamma X}) = (10 m_4, \ 2 \times 10^{-3})$]
{\includegraphics[width=0.32\textwidth]{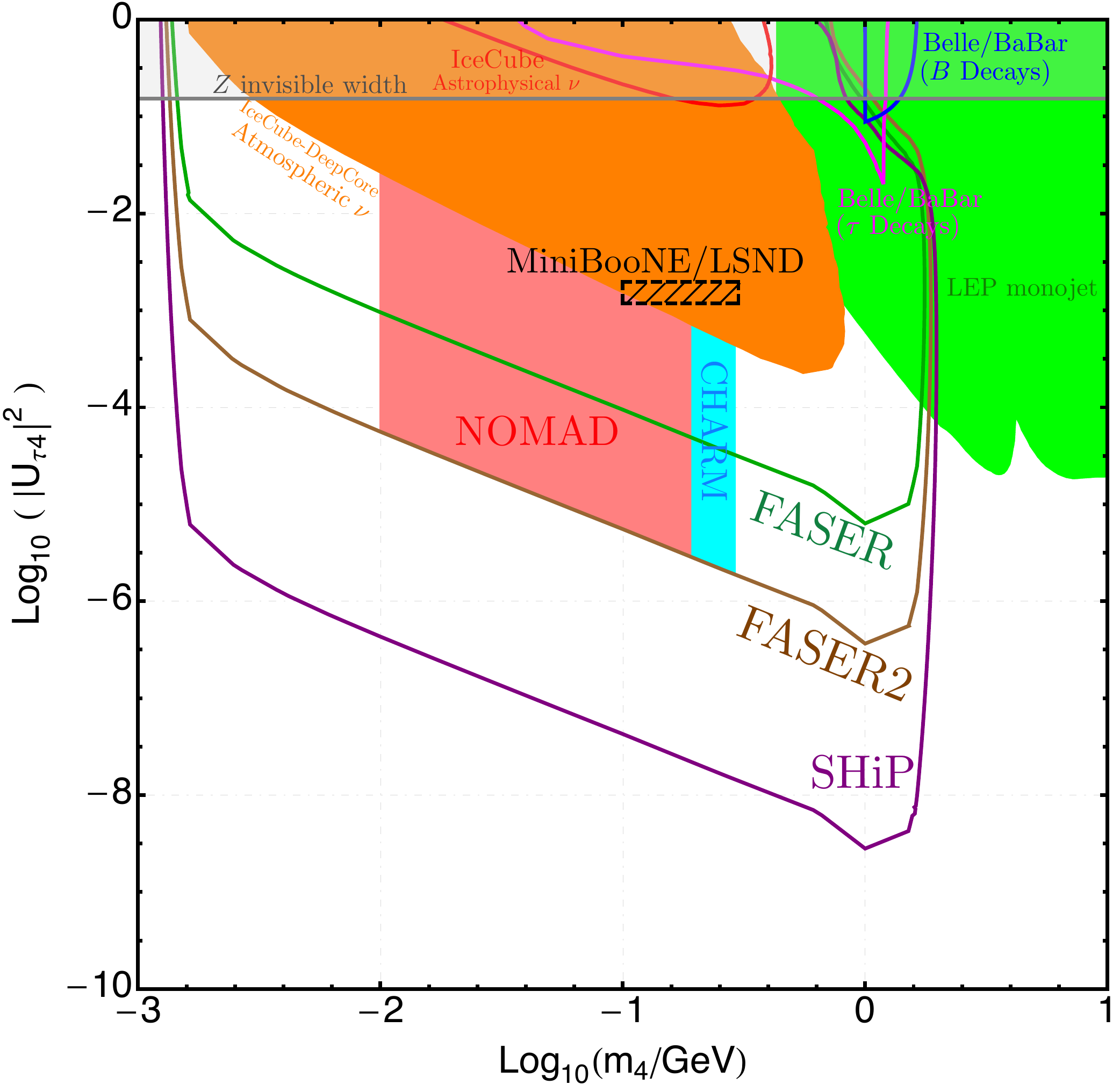}\label{fig_BP6_modelB}} \,\,
  \caption{(Model B) Current constraints on the (tau) active-sterile mixing $| U_{\tau4}|^2$, with varying $\nu_s$-philic gauge boson parameters ($m_X, \epsilon_{\gamma X}$). The constraints from CHARM (red shaded), NOMAD (cyan shaded), IceCube-DeepCore (orange shaded), LEP monojet (green shaded) and $Z$ boson invisible width (gray shaded with horizontal gray line) are shown. Also, the future expected limits from FASER (green solid line), FASER2 (brown solid line), SHiP (purple solid line), $B\to D \tau \nu_4$ (blue solid line)/$\tau^- \to \nu_4 \pi^- \pi^+ \pi^-$ (magenta solid line) at B-factories and IceCube-Gen2 double cascade search with astrophysical neutrino (red solid line) are shown. Vertical gray solid line comes from the lower bound of $m_X$ from experimental searches. We show the favored parameter region for MiniBooNE/LSND low energy excess as the black dashed box for $m_4 < m_X$ case from Ref.~\cite{Ballett:2018ynz}.\footnote{Nevertheless, as mentioned in \cite{Ballett:2018ynz}, to obtain the low energy excess in the neutrino oscillation experiments, one also needs to have the sterile-active mixing for muon neutrinos $U_{\mu4}$ too. We leave the detailed analysis for our future work.}}\label{fig_BP_modelB}
\end{figure*}


\section{Conclusion and Discussion}
\label{sec:conclusion}

Sterile neutrinos may have their own gauge charges and interact with the 
gauge boson, $X^\mu$. 
We consider two $U(1)_X$ models in this paper, where $X=B-L$ (Model A) and $X=s$, sterile-specific (Model B) as concrete examples and study phenomenology for various experiments, some are existing and some are planned.  
Focusing on $U_{\tau 4}$ mixing between the tau neutrino and the sterile neutrino, we collect all relevant results from various collider experiments, neutrino telescope, and beam dump experiments and also provide the future sensitivities for FASER, SHiP and IceCube upgrade. Our main results are summarized in Fig. \ref{fig_BP_modelA} and Fig. \ref{fig_BP_modelB}.
In particular the rectangular shape region in Fig. \ref{fig_BP_modelB} is the preferred 
parameter space to resolve the short baseline neutrino anomaly from MinBooNE, LSND \cite{Ballett:2018ynz}.

More generic mixings, $U_{\ell 4}$ for $\ell =e, \mu, \tau$ with general flavors of active neutrinos, are certainly interesting theoretical possibilities even though we focus only on $\ell=\tau$ in this paper. Especially, for $m_4 < 1$GeV, the $(m_4, | U_{\tau4}|^2)$ parameter region of model A and B will be tested by the near future searches from the FASER and proposed experiments, SHiP. [Work in progress] 
%
%

%

\begin{acknowledgments}
We would like to thank F. Kling for providing valuable data on tau neutrino flux at FASER which is studied in Ref.~\cite{Abreu:2019yak}. 
The work is supported in part by KIAS Individual Grants, Grant No. PG021403 (PK) and 
Grant No. PG074201 (JK) at Korea Institute for Advanced Study, and by National Research Foundation of Korea (NRF) Grant No. NRF-2019R1A2C3005009 (PK) and NRF-2018R1A4A1025334, NRF-2019R1A2C1089334 (SCP), funded by the Korea government (MSIT).
\end{acknowledgments}

\bibliographystyle{utphys}
\bibliography{biblio}

\end{document}